\documentclass[3p,times,authoryear,12pt]{elsarticle}
\usepackage{hyperref}
 \usepackage{verbatim}  
\usepackage{ecrc}
\volume{00}
\journalname{New Astronomy}
\runauth{Tian et al.}
\jid{NA}
\firstpage{1}
\jnltitlelogo{New Astronomy}
\CopyrightLine{2011}{Published by Elsevier Ltd.}

\usepackage{amssymb}
 \usepackage{amsthm}
\usepackage{epstopdf} 
\usepackage{array}
\usepackage{graphicx}
\usepackage{multirow}
\usepackage{arydshln}
\usepackage{threeparttable}
\usepackage{booktabs}
\usepackage{latexsym}

\newcommand{\aap}{    {Astron. Astrophys. }}

\newcommand{\apj}{    {Astrophys. J. }}
\newcommand{\apjl}{   {Astrophys. J. Lett. }}
\newcommand{\apjs}{   {Astrophys. Suppl. }}
\newcommand{\apss}{   {Astrophys. Space Sci. }}

\newcommand{\jgr}{    {J. Geophys. Res. }}

\newcommand{\nat}{    {Nature }}

\newcommand{\solphys}{{Solar Phys. }}

\newcommand{\na}{     {New Astronomy }}

\newcommand{\nfig}[1]{Figure~\ref{#1}}
\newcommand{\speed}[1]{#1 km~s${}^{-1}$}

\begin{document}

\begin{frontmatter}
\title{Formation and Eruption of a Double-decker Filament Triggered by Micro-bursts and Recurrent Jets in the Filament Channel}

\author[a,c]{Zhanjun Tian}

\author[a,b,d]{Yuandeng Shen}
\ead{ydshen@ynao.ac.cn}

\author[a,d]{Yu Liu}

\address[a]{Yunnan Observatories, Chinese Academy of Sciences, Kunming, China, 650216}
\address[b]{State Key Laboratory of Space Weather, Chinese Academy of Sciences, Beijing 100190, China}
\address[c]{University of Chinese Academy of Sciences, Beijing, China,100049}
\address[d]{Center for Astronomical Mega-Science, Chinese Academy of Sciences, Beijing, 100012, China}

\begin{abstract}
We present the observations of a double-decker filament to study its formation, triggering, and eruption physics. It is observed that the double-decker filament was formed by splitting of an original single filament. During the splitting process, intermittent bright point bursts are observed in the filament channel, which resulted in the generation of the upper filament branch. The eruption of the newly formed double-decker filament was possibly triggered by two recurrent two-sided loop jets in the filament channel and the continuous mass unloading from the upper filament body. The interaction between the first jet and the filament directly resulted in the unstable of the lower branch and the fast rising phase of the upper branch. The second jet occurred at the same site about three hours after the first one, which further disturbed and accelerated the rising of the lower filament branch. It is interesting that the rising lower branch overtook the upper one, and then the two branches probably merged into one filament. Finally, the whole filament erupted violently and caused a large-scale coronal mass ejection, leaving behind a pair of flare ribbons and two dimming regions on the both sides of the filament channel. We think that the intermittent bursts may directly result in the rearrangement of the filament magnetic field and therefore the formation of the double-decker filament, then the recurrent jets further caused the fully eruption of the entire filament system. The study provides convincing evidence for supporting the scenario that a double-decker filament can be formed by splitting a single filament into two branches.
\end{abstract}


\end{frontmatter}

\section{Introduction}
Solar prominences are thread-like clouds consisting of relatively cool (5000-8000K) and dense (10$^{10}$-10$^{11}$cm$^{-3}$) magnetized plasma suspended in the hot tenuous corona \citep[e.g.,][]{hiraya85,foukal04}. Prominences and filaments are the same entities with the former over the limb and the latter against the solar disk. In the following, we use the term ``filament'' throughout the paper. Filaments are always observed along photospheric magnetic polarity inversion lines (PIL) that denote filament channels. The magnetic field of filament channels are approximately aligned with PILs, which may {\bf be} caused by the photospheric activities such as supergranular shearing, convecting motions, and flux cancellations \citep{wang07}. In extreme ultraviolet (EUV) observations, filament channels are darker than their surroundings. Besides the main spine along the filament channel, a filament often hosts a few lateral feet or barbs. In high resolution observations, a typical filament shows more fine structures such as bubbles and ubiquitous plasma flows in every parts of the filaments \citep{berger08,berger10,shen15}.

Although it has been widely accepted that filaments are supported by coronal magnetic fields against gravity, the structure and formation process of the magnetic fields in and around filaments remains unclear. Different magnetic topologies have been proposed to understand the magnetic structures of filaments, such as the sheared arcade \citep[e.g.,][]{anti94}, and/or flux rope \citep[e.g.,][]{amari00} models. So far, the formation mechanism of filaments are still unclear. Various scenarios have been proposed to explain the formation of filaments. For example, filaments can be formed by emerging from the photosphere \citep[e.g.,][]{2009ApJ...697.1529F, 2008SoPh..247..103Y}, rearranging of an existing magnetic structure due to photospheric flux cancellation and/or coronal reconnection \citep[e.g.,][]{2015ApJ...803...86Y,2016ApJ...830...16Y,2016ApJ...816...41Y,yan15}.

The eruption of filaments are the most striking phenomena in the solar atmosphere, and they can cause devastating influence on the near-Earth space environment, especially associated with solar flares or/and coronal mass ejections (CMEs) \citep[e.g.,][]{lin03,hud06}. The eruption of filaments are often accompanied by various solar phenomena, such as flares \citep[e.g.,][]{2003ApJ...593L.137L, 2011RAA....11..594S,2012ApJ...745....9Y,2015ApJ...805...48B}, CMEs \citep[e.g.,][]{2012ApJ...750...12S,2012SoPh..279..115Y}, coronal dimmings \citep[e.g.,][]{2011SoPh..270..551Y, 2011NewA...16..276B}, global EUV waves \citep[e.g.,][]{2012ApJ...754....7S, 2017ApJ...851..101S, 2013ApJ...775...39Y}.  Filaments supported by different magnetic fields have been explained with different kinds of physical mechanisms. Besides the well-known ``Lin-Forbes'' model \citep[][]{lin00}, ``tether-cutting" model \citep[e.g.,][]{2001ApJ...552..833M, 2016ApJ...818L..27C}, and ``breakout" model \citep[e.g.,][]{1999ApJ...510..485A, 2012ApJ...750...12S, 2016ApJ...820L..37C}, there are still many other theoretical filament eruption models based on the loss of equilibrium. For example, filament eruptions can be formed directly from slow evolution of the magnetic fields leading to ``catastrophic loss of equilibrium" in the relevant magnetic structures \citep[e.g.,][]{forbes90,lin00,lin03,lin04,low01}. In addition, the helical kink instability \citep[e.g.,][]{toeroe04,liu09}, the torus instability \citep[e.g.,][]{kliem06,schr08}, and the mass-unloading mechanism are also important for destabilizing filaments \citep[e.g.,][]{low03,bi14,2014NewA...26...23X,qin17,jenkin18}.

Recently, \citet{liu12} reported so-called ``double-decker filament'', which is composed by two filaments in the same filament channel but separated in height. The two filament components are named as upper branch (UB) and lower branch (LB), respectively. The authors proposed two possible scenarios for explaining the formation of the double-decker filaments. One is the emerge filament model in which a new filament emerges below the old one to compose a double-decker filament. The other is the splitting model in which a single filament splits into two filaments for internal reconnection. This model is motivated by partial eruption proposed by \cite{gil01}. In the partial eruption scenario, the reconnection within a single stretched flux rope or flux bundles vertically splits along the original rope axis into two ropes with the same handedness. Furthermore, the splitting often occurs before or during the eruption in these cases, which means that the two new branches can not be survival for much longer time. \cite{kliem14} further demonstrated that such a special filament configuration can be stable if an external toroidal (shear) field component exceeding a threshold value. If the external toroidal field component decreases sufficiently, then both filaments turned unstable. In addition, the authors also suggested that the transfer of magnetic flux and current to the UB is the key mechanism responsible for the loss of equilibrium for double-decker filaments and this phenomenon was also discovered by other researchers \citep[e.g.,][]{zhu14,zhang14}. However, the authors did not provide any believable observational evidence for supporting their models.

In the present paper, we present unambiguous evidence to support the splitting model of double-decker filaments. The event occurred on 2017 November 16 in the quiet-Sun region on the south of NOAA active region AR12687 (S07E43), in which a quiescent filament first split into two branches due to some small-scale explosive bursts in the filament channel, then it reached a new equilibrium state and lasted for a few hours. During this time interval, two recurrent two-sided loop jets in the filament channel and mass unloading from the UB are observed, which are thought to be the trigger agents corresponding to the eruption of the newly formed double-decker filament. The used observations are introduced in Section 2, results are given in Section 3, conclusions and discussion are presented in the last section.

\section{Observations } 
The main data used in this study were taken by the Atmospheric Imaging Assembly \citep[AIA;][]{lemen12} and the Helioseismic and Magnetic Imager \citep[HMI;][]{schou12} on board the {\em Solar Dynamics Observatory} \citep[{SDO};][]{pesnel12}. The AIA provides full-disk EUV and ultraviolet (UV)-visible images in ten channels. The pixel size of the AIA images is of $0.6''$, and the cadences of the EUV and UV-visible images are 12 and 24 s, respectively. The HMI provides continuous full-disk line-of-sight (LOS) magnetograms that have a cadence of 45 s and a measuring precision of 10 Gauss. The CME associated with the filament eruption is studied by using the white-light images taken by the Large Angle and Spectrometric Coronagraphs \citep[{LASCO};][]{bru95} on board the {\em Solar and Heliospheric Observatory} ({\em SOHO}).

\section{Results}  
An overview of the pre-eruption condition of the event is shown in \nfig{fo} with the AIA 304 \AA\ and the HMI LOS magnetograms. It can be seen that the filament resided in a  filament channel that was in the northwest-southeast direction. The outline of the filament is overlaid in the HMI magnetogram, which clearly shows that the filament is a quiescent filament that located in the quiet-Sun region, and the magnetic fields on both sides of the filament are of opposite magnetic polarities (see \nfig{fo} (b)). The close up view of the small region located in the filament channel is plotted in panel (c) of \nfig{fo}, which shows a small bipolar magnetic  region, in which obvious flux cancellation is detected between the positive and negative polarities. This special region was also the source region of some transient bursts and the subsequent two recurrent two-sided loop jets. The quiescent filament can be best seen in the AIA 304 \AA\ images, which with its eastern part close to the disk limb and the western part on the disk  (see \nfig{fo} (d)). The filament was split into two branches  (see \nfig{fo} (f)), i.e., the so-called double-decker filament. The newly formed double-decker filament erupted at about 18:00:00 UT, which resulted in a pair of flare ribbons and double dimming regions on the solar surface, and a large-scale CME in the outer corona. The detail evolution processes for the formation, triggering, and eruption of the filament are described in following sections.

\subsection{ Formation of the Double-decker Filament}
\begin{figure}    
\centerline{\includegraphics[width=1\columnwidth,clip=]{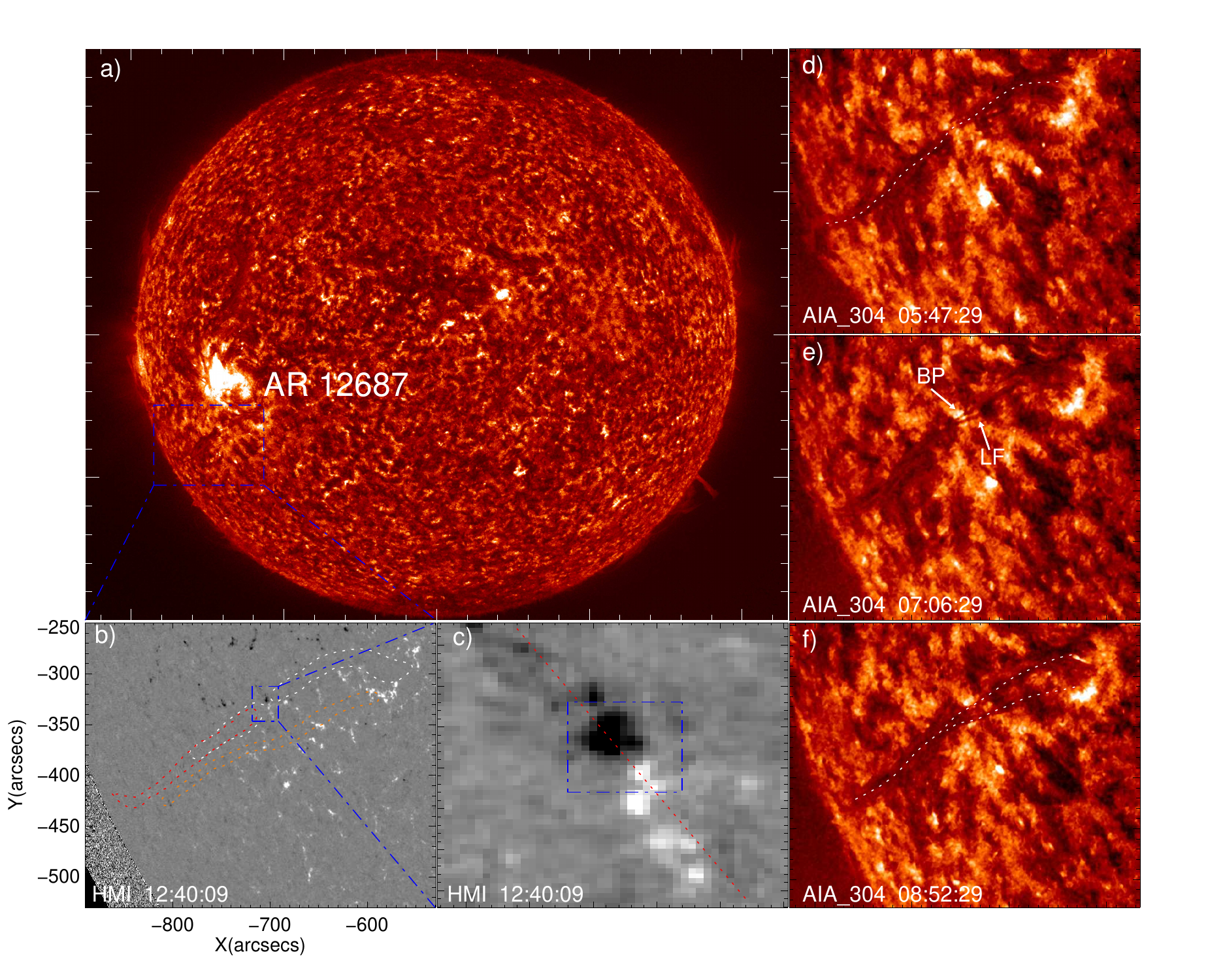}}
\caption{Panel (a) is a full-disk AIA 304 \AA\ image showing the location of the filament, in which the blue box indicates the field of view (FOV) of panel (b). Panel (b) is a LOS magnetogram overlaid with the contours of the filament channel (white), the upper branch (orange) and the east section of lower branch (red), respectively. The blue box indicates the source region of two jets and the FOV of panel (c). Pannel (c) shows the closed up view of the eruption source region of the jets. The black and white patches in the magnetograms represent the negative and positive magnetic field regions, respectively. Panels (d) -- (f) show the formation and separation of the upper branch. The white dotted line in panel (d) and (f) indicate the spine of filaments. ``BP" and ``LF" in panel (e) indicate the bright point and loop-like upper branch, respectively. Animations (animation1.mpg, animation2.mpg) of the AIA 304 \AA\ and HMI LOS magnetogram observations are available in the online journal.}
\label{fo}
\end{figure}

To study the formation process of the double-decker filament, we trace back the filament to at about 05:00:00 UT, that is about 13 hours before the start of the violent filament eruption. At 05:47:29 UT, it is obvious that the filament was a single long dark feature in the filament channel (see the dotted line in \nfig{fo} (d)). During the following several hours, some intermittent bright point bursts are observed around the bipolar magnetic region in the filament channel. The arrow labelled ``BP'' in panel (e) of \nfig{fo} points to a typical bright point burst. It is observed that the filament started to split into two branches right after an obvious bright point burst in the bipolar magnetic region at about 07:06:29 UT. One can see that the filament kept stable at the original location, a loop-like filament thread started to separate from the west section ( the right of ``BP") of the original filament and showed a southward slow rising motion (see the arrow labelled ``LF'' in panel (e) of \nfig{fo}). Here, we define the stable part of the filament as the LB, while the rising loop-like part as the UB. The rising motion of the UB made the separation distance to the LB increasing in time and the separation slowly slipped to the east section. About two hours later, it is easy to distinguish the two branches in the AIA 304 \AA\ images, due to the increased separation distance between them (see \nfig{fo} (f)). The positions of the UB and LB at the time of 12:40:29 UT (see panel (a) in \nfig{jets} ) are all overlaid on the HMI LOS magnetogram in \nfig{fo} (b), when the separation was well developed. One can see that the {\bf projection position of} UB located on the positive magnetic region on the {\bf southern} side of the filament channel (see the orange contour in \nfig{fo} (b)), which suggests that the UB had rose to a higher altitude. The slow rising motion of the east section of UB lasted for about 3 hours with a projected speed of about \speed{0.68}. For more details about the formation of the double-decker filament and the evolution of the magnetic bipolar, one can see the online animations (animation1.mpg, animation2.mpg) associated with \nfig{fo}.

The evolution pattern described above suggests that the double-decker filament was formed by splitting the original single filament, due to some intermittent bright point bursts in the filament channel. The intermittent bright point bursts may represent some kind of magnetic reconnections among field lines that rooted in the source region, where the positive- and negative-polarity flux meet together.  Those magnetic reconnections can cause the rearrangement of the magnetic field distribution in the filament and its surroundings \citep[e.g.,][]{wang07,wang13}. More discussions about the relations between the two phenomena, bursts and splitting, are displayed in Section \ref{c_d}.

\subsection{Triggering and Eruption of the Filament}
\begin{figure}    
\centerline{\includegraphics[width=1\columnwidth,clip=]{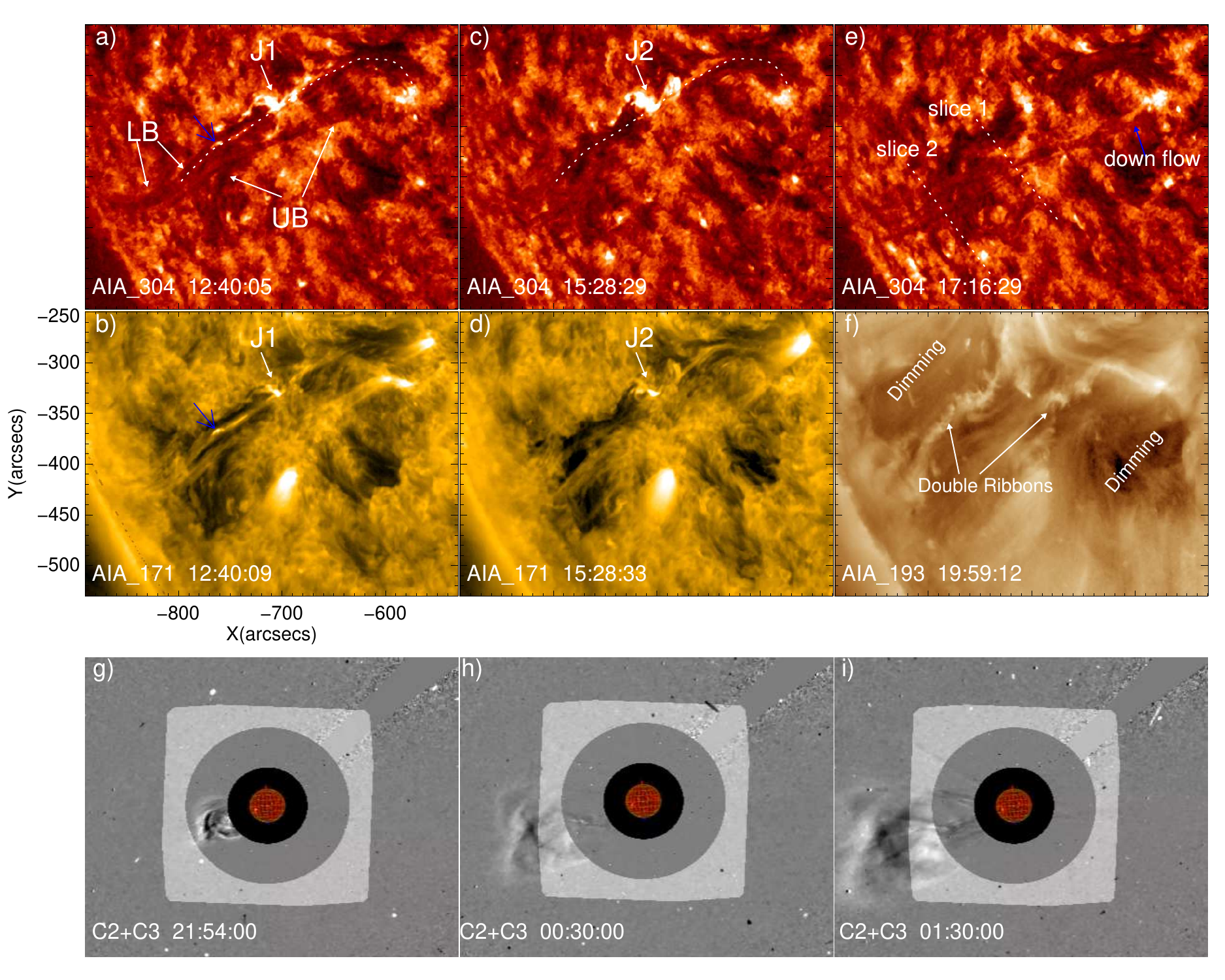}}
\caption{The top, middle, and bottom rows show the AIA 304 \AA\, AIA 171 \AA\, and composited LASCO images, respectively. Letters ``J1'', ``J2'' indicate the first and the second two-sided loop jets, while ``LB'' and ``UB'' indicate the lower and upper filament branches, respectively. The arrow in panel (e) points to the drainage mass. The pair of flare ribbons and the double dimming regions are indicated in panel (f). In the composited images, the inner red part indicates the solar disk with the AIA 304 \AA\ images, while the gray middle and outer are the LASCO C2 and C3 running difference images, respectively. An animation (animation2.mpg) made from AIA 304 \AA\ and 171 \AA\ observations is available in the online journal.}
\label{jets}
\end{figure}

At about 12:30:00 UT, a two-sided loop jet \citep[e.g.,][]{alex99,tian17,ning16} occurred at the bipolar magnetic region in the filament channel, whose arms were in opposite directions along the filament channel (see the arrow labelled with ``J1'' in \nfig{jets} (a) and (b)). The eastward and westward arms of the jet directly interacted with LB, and bright plasma blobs are observed propagating along the filament threads (see \nfig{jets} (a) and (b)). The interaction of the jet and the LB may directly inject plasma into the filament, which can further cause the unstable of LB due to the increased mass with momentum \citep{guoj10}. Additionally, this is similar to the study performed by \citep{2005ApJ...631L..93L}, where the authors observed that filaments are generated by the injection of recurrent jets, and suggested that there should be a direct link between the filament axial fields and the large-scale background fields along which the ejecta can be driven into the filament. It is interesting that the UB experienced a faster rising phase relative to the previous splitting stage right after the start time of the jet, which suggests that the jet not only disturbed the LB but also the dynamics of the entire filament system.

It is noted that the rising of UB stopped at a plateau height at about 14:00:00 UT and lasted for about three hours until the start of the fast eruption of the filament at about 17:00:00 UT. During this time interval, intermittent mass drainage is observed along the western leg of UB (see the blue arrow in \nfig{jets} (e)), which might be important for the loss of equilibrium of the filament \citep[e.g.,][]{low03,bi14,qin17,jenkin18}. At about 15:20:00 UT, the second two-sided loop jet occurred at the same site as the first one, and the morphology of the jet is shown with the AIA 171 \AA\ and 304 \AA\ images (see the arrow labelled with ``J2'' in \nfig{jets} (c) and (d)). In contrast to the first jet, the eastward arm resembles a loop-like feature, and it was suppressed within a short distance, while the westward arm ejected to a large distance up to the UB's west end. Due to the disturb caused by the interaction between the second jet and LB, the east section of LB started to rise in the south {\bf projection} direction and cut off from the west section of LB  which remained in the original height. For brevity, we refer the east section of the lower branch as ``LB"  in the following. It is observed that the rising speed of LB was faster than that of the UB, which caused the interaction and probably merging of the two branches at about 16:00:00 UT. Such an interaction and merging between two filaments have also been reported in previous studies \citep[e.g.,][]{su07,2017ApJ...838..131Y}. However, due to the lack of stereoscopic observations, we are unable to distinguish the merging of the two branches were true or not, since the projection effect can also cause this in observations if the two separated branches were all moving in the line-of-sight. About one hour after the interaction of the two branches, the newly merged filament started the fast rising and eruption phases at about 17:00:00 UT. During the eruption, continuous mass drainage with many bright blobs are observed along the western leg of the erupting filament, which further reinforced the acceleration of the slow rising filament. The acceleration of the filament lasted for about one hour, and finally the entire filament lost equilibrium and erupted totally at about 18:00:00 UT. For more details about the two-sided loop jets and their interaction to the filament, one can see the online animation (animation3.mpg) associated with \nfig{jets}.

During and after the eruption of the filament, a pair of bright flare ribbons are observed on the both sides of the filament channel. In addition, two obvious dimming regions appeared at the outer sides of the two flare ribbons (see \nfig{jets} (f)). These low coronal phenomena are frequently observed in a large number of studies, and they are thought to be the typical eruption characteristics accompanying with filament eruptions. The present filament eruption further resulted in a large-scale CME in the outer corona, which was recorded by LASCO C2 and C3 coronagraphs. The morphology of the CME is displayed in \nfig{jets} (g) -- (i) with the composited images made from AIA 304 \AA\ (inner), LASCO C2 (middle), and LASCO C3 (outer) observations. To better show the CME, we use the running difference images of LASCO C2 and C3 in the composited images. Here, a running difference image is obtained by subtracting a image by the previous one in time, in which moving features can be observed more clearly. The CME was at an average speed of \speed{610} within the FOV of C2 and C3, and its first appearance times in the FOVs of LASCO C2 and C3 were at 20:18:00 UT and 22:06:07 UT, respectively.

\begin{figure}    
\centerline{\includegraphics[width=1\columnwidth,clip=]{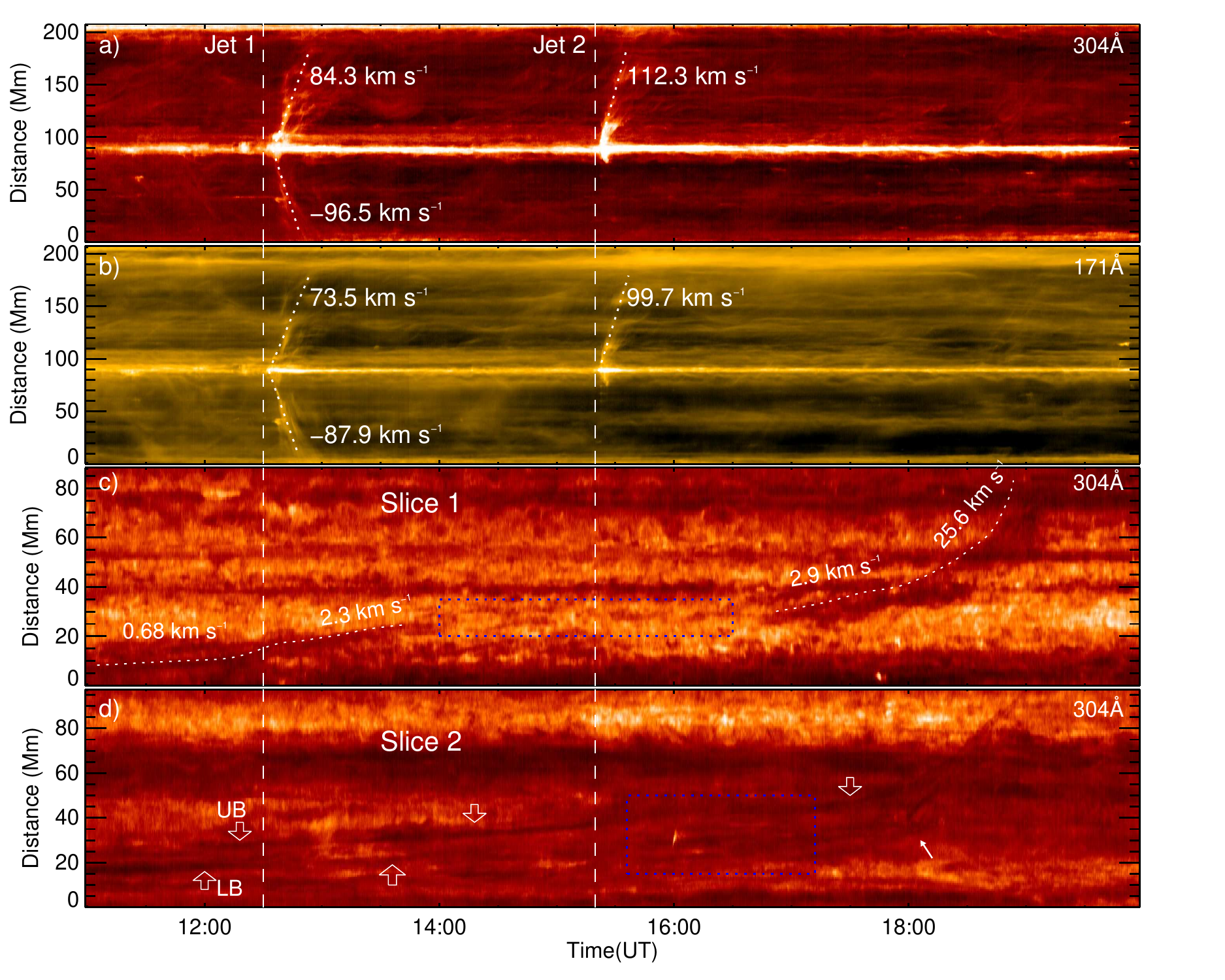}}
\caption{Time-slice diagrams show the kinematics of the recurrent jets and the evolution of the double-decker filament from 11:00:05 UT to 19:59:17 UT. Panel (a) and (b) are made from AIA 304 \AA\ and 171 \AA\ images along the filament channel as shown by the dotted curve in \nfig{fo} (d). Panels (c) and (d) are made from AIA 304 \AA\ images along the two lines across the filament axis as shown in \nfig{jets} (e). The dotted lines are linear fit to the paths of the jets, and the speeds are also plotted in the figure. In panel (c), The white dotted lines are the quadratic fit to the rising UB, and the average speeds are also labeled. The blue box indicates the time range time interval of the new equilibrium of the UB. In panel (d), the white down (up) hollow arrows indicate the upper (lower) filament branch, respectively. The blue box indicates the time interval during which interaction and emerging of the two branches occurred. Meanwhile, the rapid lifting of LB at about 18:00:00UT is marked by a white arrow. The two vertical dotted lines indicate the start times of two jets, respectively.}
\label{td}
\end{figure}

The entire evolution process of the filament is also studied by using time-distance diagrams. To obtain a time-distance diagram, we first obtain the one-dimensional intensity profiles along a specified path at different times, and then a two-dimensional time-distance diagram can be generated by stacking the obtained one-dimensional intensity profiles in time.  To study the kinematics of the two recurrent two-sided loop jets, Time-distance diagrams are made from AIA 304 and 171 \AA\ observations along the filament axis (see panels (a) and (b) in \nfig{td}), respectively. The ejection speed of the two jets are measured by applying a linear fit to the inclined bright features that represent the arms of the jets. The measurement results indicate that the speeds of the eastward (westward) arms of the first jet is about \speed{-96.5 (84.3)} in the 304 \AA\ time-distance diagram, while the speed of the westward arm of the second jet is about \speed{112.3}. The eastward arm of the second jet was confined within a short distance, thus that we do not measure its speed. In the 171 \AA\ time-distance diagram, the corresponding ejection speeds of the jets are relatively slower than those obtained in the 304 \AA\ time-distance diagram.

Panels (c) and (d) in \nfig{td} are time-distance diagrams made from AIA 304 \AA\ images along the two dotted lines across the filament axis (see \nfig{jets} (e)), respectively. These time-distance diagrams show the time evolution of the filament more clearly. The time-distance diagram in \nfig{td} (c) shows the evolution of the UB better. One can see that the filament underwent a slow rising phase with a speed of about \speed{0.68}. Right after the occurrence of the first two-sided loop jet, the UB started a faster constant rising phase with a speed of about \speed{2.3}. The change of the rising speed obviously indicates the influence resulted from the first jet. Interestingly, the western part of UB stopped to rise during the time interval from about 14:00:00 UT to 16:30:00 UT (see the blue box in \nfig{td} (c)), which suggests that the rising UB reached a new equilibrium state. In addition, the mass drainage from the filament can also be observed in the time-distance diagram during this time interval. Finally, the filament started its violent eruption process after 17:00:00 UT. It is clearly that the eruption of the filament underwent a slow and a following fast eruption phases, and the corresponding average speeds are about \speed{2.9 and 25.6}, respectively.

The time-distance diagram in \nfig{td} (d) shows the evolution of the both branches and their interaction process clearly, in which the LB and UB are indicated with arrows labelled with letters ``LB'' and ``UB'', respectively. It can be seen that the LB kept stable before the first jet, then it showed a little rise during the time interval of the two jets. After the second jet, the LB underwent a fast rising phase and therefore led to the interaction and merging of the two filament branches during the time interval from 15:30:00 UT to 17:00:00 UT. The fast rising of the LB manifested the influence caused by the second jet. The merged filament started its eruption process after 17:00:00 UT, and the two acceleration phases are also clear in this time-distance diagram.

\subsection{Magnetic Flux Variation}
\begin{figure}    
\centerline{\includegraphics[width=1\columnwidth,clip=]{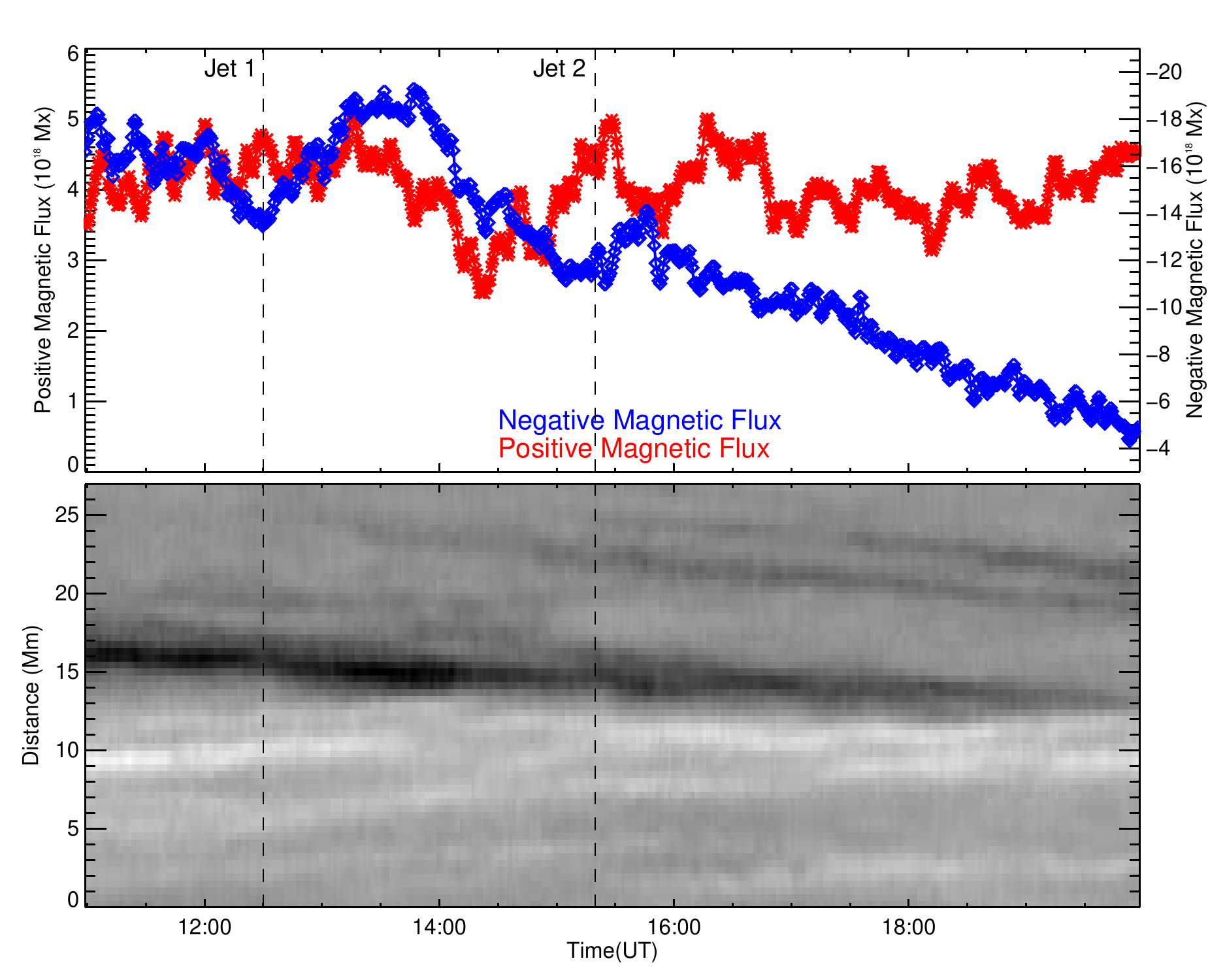}}
\caption{Panel (a) shows the variations of the positive (red) and negative (blue) magnetic fluxes in time from 10:58:55 UT to 19:58:09 UT, which are measured within the blue box region as shown in \nfig{fo} (c). Panel (b) shows the magnetic field evolution along the line as shown in \nfig{fo} (c). The two vertical dotted lines indicate the start times of two jets. }
\label{flux}
\end{figure}

The magnetic flux variation around the eruption source region of the two-sided loop jets are studied in detail, and the results are plotted in \nfig{flux}. The positive and negative magnetic fluxes within the blue box region as shown in \nfig{fo} (c) are plotted as red and blue curves in \nfig{flux} (a), respectively. A time-distance diagram (\nfig{flux} (b)) along the red dashed line as shown in \nfig{fo} (c) is also generated to show the spatial evolution of the positive and negative magnetic polarities in the eruption source region of the jets. As shown in \nfig{flux} (a), the positive magnetic flux displayed a moderate and irregular fluctuation at a level of about $4 \times 10^{18}$ Mx during the observation time interval of our study. Whereas the magnitude of negative flux showed a continuous decreasing trend from $16 \times 10^{18}$ to $4 \times 10^{18}$ Mx except for the short period of time after first jet, corresponding to an average loss rate of $\backsim 2 \times 10^{18}$ Mx hr$^{-1}$, which is consistent with the results reported in  \citet{wang13}. 

Here, we focus on the two short time intervals around the beginnings of the two jets to analyze the causal relationship between the magnetic flux variations and the observed coronal jets. The start times of the jets are marked in \nfig{flux} with two vertical dashed lines. As one can see that the positive (negative) flux before the start time of the first jet showed a rapid increase (decrease) period of about half an hour, then it changed to decrease (increase) trend right after the start time of the jet. The variation of the fluxes around the start time of the second jet showed a similar evolution pattern. These variation trends of the magnetic fluxes suggest the emergence (cancellation) of the positive (negative) magnetic flux before the start of the jets. However, right after the start of the jets the variation trend of the positive (negative) flux changed into cancellation (emergence). In the time-distance diagram (\nfig{flux} (b)), the convergence motion between the positive and negative magnetic polarities is obviously during the entire observation time interval. The cancellation and emergence of the magnetic fluxes can also be clearly identified. Therefore, the eruption of the jets were possibly triggered by the alternating emergence and cancellation of the positive and negative magnetic fluxes.

\section{Conclussions \& Discussions }
\label{c_d}
We present the observational analysis of a double-decker filament close NOAA active region AR12687 on 2017 November 16, by using the high temporal and spatial resolution data taken by the {\em SDO}. For the first time, we report the detailed evolution of the double-decker filament, including the formation, triggering, and eruption processes. This study provides a convincing example for supporting the prediction that a double-decker filament can be formed by splitting an original single filament into two filament branches.

The observational results indicate that the formation of the double-decker filament in the present case was due to the splitting of an original single filament. This physical picture is consistent with one of the possible formation mechanisms proposed by \citet{liu12}. At the very beginning, it is observed that a quiescent filament resided in a filament channel about 05:00:00 UT. In the following several hours, some intermittent bright point bursts are observed around the small bipolar magnetic regions in the filament channel. Simultaneously, the paired opposite polarities in the eruption source region showed obvious converging motion and magnetic cancellation between them. The bright point bursts represent energy releasing events attributed to magnetic reconnection, which can convert short  field lines into upward longer and downward shorter loops \citep{van89}. So we can image that the short filament threads rooted in the positive and negative polarities in the eruption source region could be converted to longer filament threads via magnetic reconnection \citep{tian17}. The filament underwent a long splitting process of about 6 hours, in which the LB remained at the original height while the loop-like UB showed obvious slow rising movement to the south {\bf projection} direction at a speed of about \speed{0.68}. For the splitting of a single filament into two branches, the model proposed in \citet{liu12} did not discuss the detailed physical process. \cite{gil01} proposed a model in which the reconnection occurs inside the filament, this can cause the splitting of the filament. Another possibility for the filament splitting could be caused by mixed kink instability and magnetic reconnection process in the bipolar, which can effectively transfer and rearrange the magnetic field distribution in the filament. The mixed kink instability are proposed by \citet{mei18} in their parametric study on kink instabilities. According to the linear theory of MHD instabilities for cylindrical plasma columns \citep{goedbloed10}, complex twist turns distribution inside the realistic filament may allow appearance of external and internal kink instabilities in the meanwhile, and the mixed kink refer to this situation. The resultant evolution of filaments demonstrate some interesting and complicated behaviors, specifically, the splitting process of filaments. Since we did not observed reconnection characteristics in the filament body, the splitting of the present filament was not suitable for the model proposed by \cite{gil01}. We think that the splitting of the filament in the present case seems more suitable for the mechanism of mixed kink instability proposed by \citet{mei18}, in which the intermittent bright point bursts underneath the filament may provide successive disturbances to the filament system and therefore trigger or accelerate the external and internal kink instabilities of the filament.

The triggering of the filament eruption is analyzed in detail. It is found that the triggering of the filament eruption was tightly associated with the eruption of two recurrent two-sided loop jets originated from the small bipolar region. The variations of the magnetic fluxes suggest that the recurrent jets are possibly caused by alternating flux emergence and cancellation activities in the bipolar region. It is observed that the first jet directly interacted with the LB, and right after the beginning of the jet the UB was accelerated to a faster rising phase at a speed of \speed{2.3}. The second jet occurred at about 15:20:00 UT and interacted with the LB again, which directly resulted in the rising of the LB. Obviously, the rising of the LB was faster than the upper one. Therefore, the two filament branches interacted and probably merged during the time interval from 15:30:00 UT to 17:00:00 UT, during which substantial mass drainage from the UB is also observed. Finally, the filament erupted totally at about 17:00:00 UT, which caused a large-scale CME at an average speed of \speed{610}. In addition, low coronal features such as the  pair of flare ribbons and the double dimming regions are also observed to confirm the eruption of the filament.

Based on the observational results, the temporal and spatial relationship between the recurrent jets and the onset of the filament  eruption suggest that the quasi-equilibrium state of the newly formed double-decker filament was destroyed by the jet activities and the mass unloading of the UB. The recurrent two-sided loop jets can not only provide plasma  material and momentum to the filament, but also can rearrange the filament magnetic field environment. On the other hand, the continuous mass unloading of the UB removed a part of weight from the filament. Consequently, the system can not keep balance between the downward gravity force and upward magnetic force. Therefore, the entire filament system will lose equilibrium and erupt when the upward force exceed the downward gravity. This is consistent with the models of mass unloading eruptions, which predict that significant material drainage may facilitate an eruption \citep{low03}. The model suggests that sufficiently large mass in filament can contribute to equilibrium. That is to say if mass removed, the magnetic environment would be free to expand and attempted to find a new equilibrium by increasing the height of the flux rope.

Solar jets are ubiquitous in the solar atmosphere, and they are often found to be associated with magnetic flux emergence and cancellation activities \citep[e.g.,][]{liu04,2012ApJ...745..164S,2017ApJ...851...67S}. Traditional theoretical studies proposed that a solar jet is caused by the magnetic reconnection between an emerging bipole and its ambient open fields \citep[e.g.,][]{1995Natur.375...42Y}. In particular cases, if an emerging magnetic bipole reconnects with the overlying horizontal field lines, the consequence should be the generation of a two-sided loop jet \citep{1995Natur.375...42Y}. In addition, two-sided loop jets can also be produced by the magnetic reconnection between adjacent filamentary threads \cite{tian17}. Recent high resolution observations indicate that many jets are dynamically associated with the eruption of mini-filaments \citep[e.g.,][]{2012ApJ...745..164S,2017ApJ...851...67S,2012NewA...17..732Y,2014ApJ...796...73H,2017ApJ...835...35H,2017ApJ...842L..20L,2018ApSS.363...26L}. In addition, it is also found that solar jets are tightly related to other large-scale solar eruption phenomena, in which solar jets often play a trigger role. For example, through interaction with other magnetic structures, jets can cause sympathetic CMEs \citep[e.g.,][]{2008ApJ...677..699J,2012ApJ...745..164S} and filament oscillations \citep[e.g.,][]{2017ApJ...851...47Z}. In particularly, \cite{2005ApJ...631L..93L} found that some filaments can be quickly formed by trapping the cold material supplied by jets. In the present study, we observed the interaction of two-sided loop jets interacted with and provided mass to the double-decker filament, which triggered the loss of equilibrium and fully eruption of the filament system by rearranging the magnetic fields. So far, such kind of observations are still very scarce, more similar observational and theoretical studies are needed to figure out the detailed physical process during the jet-filament interaction period.

In summary, by analyzing the detailed evolution of a quiescent filament of about 15 hours, the observational results reveal the formation, triggering, and eruption details. The present study provides convincing evidence for supporting the scenario that a double-decker filament can be formed by splitting an original single filament due to some small-scale transient explosive events. The triggering of the filament eruption was caused by two recurrent two-sided loop jets in the filament channel and the mass drainage from the filament body, which resulted in the disequilibrium of the filament. The eruption of filaments is the result of multiple factors. The present study provides a possible explanation, we do not exclude other interpretations. Studying the dynamics of filaments formation, stability, and destabilization can shed light on the precursor of eruptive events. The results of the present study are highly encouraging for future investigations. More observational and simulation works are needed in the future to understand the filaments dynamics in detail.

\section*{Acknowledgements}
We thank the anonymous referee for his/her constructive comments which are valuable for improving the quality paper. We also thank the excellent observations provided by the {\em SDO} and {\em SOHO} teams. This work is supported by the Natural Science Foundation of China (11773068, 11633008,11403097, 11503084), the Yunnan Science Foundation (2017FB006, 2015FB191), the Specialized Research Fund for State Key Laboratories, the Youth Innovation Promotion Association (2014047) of Chinese Academy of Sciences, and the grant associated with the Project of the Group for Innovation of Yunnan Province.

\newpage

\begin{thebibliography}{}
\expandafter\ifx\csname natexlab\endcsname\relax\def\natexlab#1{#1}\fi

\bibitem[{{Alexander} \& {Fletcher}(1999)}]{alex99}
{Alexander}, D., \& {Fletcher}, L. 1999, \solphys, 190, 167

\bibitem[{{Amari} {et~al.}(2000){Amari}, {Luciani}, {Mikic}, \&
  {Linker}}]{amari00}
{Amari}, T., {Luciani}, J.~F., {Mikic}, Z., \& {Linker}, J. 2000, \apjl, 529,
  L49

\bibitem[{{Antiochos} {et~al.}(1994){Antiochos}, {Dahlburg}, \&
  {Klimchuk}}]{anti94}
{Antiochos}, S.~K., {Dahlburg}, R.~B., \& {Klimchuk}, J.~A. 1994, \apjl, 420,
  L41

\bibitem[{{Antiochos} {et~al.}(1999){Antiochos}, {DeVore}, \&
  {Klimchuk}}]{1999ApJ...510..485A}
{Antiochos}, S.~K., {DeVore}, C.~R., \& {Klimchuk}, J.~A. 1999, \apj, 510, 485

\bibitem[{{Berger} {et~al.}(2008){Berger}, {Shine}, {Slater}, {Tarbell},
  {Title}, {Okamoto}, {Ichimoto}, {Katsukawa}, {Suematsu}, {Tsuneta}, {Lites},
  \& {Shimizu}}]{berger08}
{Berger}, T.~E., {Shine}, R.~A., {Slater}, G.~L., {et~al.} 2008, \apjl, 676,
  L89

\bibitem[{{Berger} {et~al.}(2010){Berger}, {Slater}, {Hurlburt}, {Shine},
  {Tarbell}, {Title}, {Lites}, {Okamoto}, {Ichimoto}, {Katsukawa}, {Magara},
  {Suematsu}, \& {Shimizu}}]{berger10}
{Berger}, T.~E., {Slater}, G., {Hurlburt}, N., {et~al.} 2010, \apj, 716, 1288

\bibitem[{{Bi} {et~al.}(2014){Bi}, {Jiang}, {Yang}, {Hong}, {Li}, {Yang}, \&
  {Yang}}]{bi14}
{Bi}, Y., {Jiang}, Y., {Yang}, J., {et~al.} 2014, \apj, 790, 100

\bibitem[{{Bi} {et~al.}(2015){Bi}, {Jiang}, {Yang}, {Xiang}, {Cai}, \&
  {Liu}}]{2015ApJ...805...48B}
---. 2015, \apj, 805, 48

\bibitem[{{Bi} {et~al.}(2011){Bi}, {Jiang}, {Yang}, \&
  {Zheng}}]{2011NewA...16..276B}
{Bi}, Y., {Jiang}, Y.~C., {Yang}, L.~H., \& {Zheng}, R.~S. 2011, \na, 16, 276

\bibitem[{{Brueckner} {et~al.}(1995){Brueckner}, {Howard}, {Koomen},
  {Korendyke}, {Michels}, {Moses}, {Socker}, {Dere}, {Lamy}, {Llebaria},
  {Bout}, {Schwenn}, {Simnett}, {Bedford}, \& {Eyles}}]{bru95}
{Brueckner}, G.~E., {Howard}, R.~A., {Koomen}, M.~J., {et~al.} 1995, \solphys,
  162, 357

\bibitem[{{Chen} {et~al.}(2016{\natexlab{a}}){Chen}, {Zhang}, {Li}, \&
  {Ma}}]{2016ApJ...818L..27C}
{Chen}, H., {Zhang}, J., {Li}, L., \& {Ma}, S. 2016{\natexlab{a}}, \apjl, 818,
  L27

\bibitem[{{Chen} {et~al.}(2016{\natexlab{b}}){Chen}, {Du}, {Zhao}, {Wu}, {Liu},
  {Wang}, {Ruan}, {Feng}, \& {Song}}]{2016ApJ...820L..37C}
{Chen}, Y., {Du}, G., {Zhao}, D., {et~al.} 2016{\natexlab{b}}, \apjl, 820, L37

\bibitem[{{Fan}(2009)}]{2009ApJ...697.1529F}
{Fan}, Y. 2009, \apj, 697, 1529

\bibitem[{{Forbes}(1990)}]{forbes90}
{Forbes}, T.~G. 1990, \jgr, 95, 11919

\bibitem[{{Foukal}(2004)}]{foukal04}
{Foukal}, P.~V. 2004, {Solar Astrophysics, 2nd, Revised Edition}, 480

\bibitem[{{Gilbert} {et~al.}(2001){Gilbert}, {Holzer}, \& {Burkepile}}]{gil01}
{Gilbert}, H.~R., {Holzer}, T.~E., \& {Burkepile}, J.~T. 2001, ApJ, 549, 1221

\bibitem[{{Goedbloed} {et~al.}(2010){Goedbloed}, {Keppens}, \&
  {Poedts}}]{goedbloed10}
{Goedbloed}, J.~P., {Keppens}, R., \& {Poedts}, S. 2010, {Advanced
  Magnetohydrodynamics}

\bibitem[{{Guo} {et~al.}(2010){Guo}, {Liu}, {Zhang}, {Deng}, {Lin}, \&
  {Su}}]{guoj10}
{Guo}, J., {Liu}, Y., {Zhang}, H., {et~al.} 2010, \apj, 711, 1057

\bibitem[{{Hirayama}(1985)}]{hiraya85}
{Hirayama}, T. 1985, \solphys, 100, 415

\bibitem[{{Hong} {et~al.}(2014){Hong}, {Jiang}, {Yang}, {Bi}, {Li}, {Yang}, \&
  {Yang}}]{2014ApJ...796...73H}
{Hong}, J., {Jiang}, Y., {Yang}, J., {et~al.} 2014, \apj, 796, 73

\bibitem[{{Hong} {et~al.}(2017){Hong}, {Jiang}, {Yang}, {Li}, \&
  {Xu}}]{2017ApJ...835...35H}
{Hong}, J., {Jiang}, Y., {Yang}, J., {Li}, H., \& {Xu}, Z. 2017, \apj, 835, 35

\bibitem[{{Hudson} {et~al.}(2006){Hudson}, {Wolfson}, \& {Metcalf}}]{hud06}
{Hudson}, H.~S., {Wolfson}, C.~J., \& {Metcalf}, T.~R. 2006, \solphys, 234, 79

\bibitem[{Jenkins {et~al.}(2018)Jenkins, Long, van Driel-Gesztelyi, \&
  Carlyle}]{jenkin18}
Jenkins, J.~M., Long, D.~M., van Driel-Gesztelyi, L., \& Carlyle, J. 2018,
  Solar Physics, 293, 7

\bibitem[{{Jiang} {et~al.}(2008){Jiang}, {Shen}, {Yi}, {Yang}, \&
  {Wang}}]{2008ApJ...677..699J}
{Jiang}, Y., {Shen}, Y., {Yi}, B., {Yang}, J., \& {Wang}, J. 2008, \apj, 677,
  699

\bibitem[{{Kliem} \& {T{\"o}r{\"o}k}(2006)}]{kliem06}
{Kliem}, B., \& {T{\"o}r{\"o}k}, T. 2006, Physical Review Letters, 96, 255002

\bibitem[{Kliem {et~al.}(2014)Kliem, Torok, Titov, Lionello, Linker, Liu, Liu,
  \& Wang}]{kliem14}
Kliem, B., Torok, T., Titov, V.~S., {et~al.} 2014, Astrophysical Journal, 792,
  107

\bibitem[{Lemen {et~al.}(2012)Lemen, Title, Akin, Boerner, Chou, Drake, Duncan,
  Edwards, Friedlaender, Heyman, Hurlburt, Katz, Kushner, Levay, Lindgren,
  Mathur, McFeaters, Mitchell, Rehse, Schrijver, Springer, Stern, Tarbell,
  Wuelser, Wolfson, Yanari, Bookbinder, Cheimets, Caldwell, Deluca, Gates,
  Golub, Park, Podgorski, Bush, Scherrer, Gummin, Smith, Auker, Jerram, Pool,
  Soufli, Windt, Beardsley, Clapp, Lang, \& Waltham}]{lemen12}
Lemen, J.~R., Title, A.~M., Akin, D.~J., {et~al.} 2012, Solar Physics, 275, 17

\bibitem[{{Li} {et~al.}(2018){Li}, {Yang}, {Jiang}, {Bi}, {Qu}, \&
  {Chen}}]{2018ApSS.363...26L}
{Li}, H., {Yang}, J., {Jiang}, Y., {et~al.} 2018, \apss, 363, 26

\bibitem[{{Li} {et~al.}(2017){Li}, {Jiang}, {Yang}, {Qu}, {Yang}, {Xu}, {Bi},
  {Hong}, \& {Chen}}]{2017ApJ...842L..20L}
{Li}, H., {Jiang}, Y., {Yang}, J., {et~al.} 2017, \apjl, 842, L20

\bibitem[{Lin(2004)}]{lin04}
Lin, J. 2004, Solar Physics, 219, 169

\bibitem[{{Lin} \& {Forbes}(2000)}]{lin00}
{Lin}, J., \& {Forbes}, T.~G. 2000, \jgr, 105, 2375

\bibitem[{Lin {et~al.}(2003)Lin, Soon, \& Baliunas}]{lin03}
Lin, J., Soon, W., \& Baliunas, S.~L. 2003, New Astronomy Reviews, 47, 53

\bibitem[{Liu {et~al.}(2012)Liu, Kliem, Torok, Liu, Titov, Lionello, Linker, \&
  Wang}]{liu12}
Liu, R., Kliem, B., Torok, T., {et~al.} 2012, Astrophysical Journal, 756, 59

\bibitem[{{Liu} {et~al.}(2003){Liu}, {Jiang}, {Ji}, {Zhang}, \&
  {Wang}}]{2003ApJ...593L.137L}
{Liu}, Y., {Jiang}, Y., {Ji}, H., {Zhang}, H., \& {Wang}, H. 2003, \apjl, 593,
  L137

\bibitem[{Liu \& Kurokawa(2004)}]{liu04}
Liu, Y., \& Kurokawa, H. 2004, Astrophysical Journal, 610, 1136

\bibitem[{{Liu} {et~al.}(2005){Liu}, {Kurokawa}, \&
  {Shibata}}]{2005ApJ...631L..93L}
{Liu}, Y., {Kurokawa}, H., \& {Shibata}, K. 2005, \apjl, 631, L93

\bibitem[{Liu {et~al.}(2009)Liu, Su, Xu, Lin, Shibata, \& Kurokawa}]{liu09}
Liu, Y., Su, J., Xu, Z., {et~al.} 2009, The Astrophysical Journal Letters, 696,
  L70

\bibitem[{{Low}(2001)}]{low01}
{Low}, B.~C. 2001, \jgr, 106, 25141

\bibitem[{{Low} {et~al.}(2003){Low}, {Fong}, \& {Fan}}]{low03}
{Low}, B.~C., {Fong}, B., \& {Fan}, Y. 2003, \apj, 594, 1060

\bibitem[{{Mei} {et~al.}(2018){Mei}, {Keppens}, {Roussev}, \& {Lin}}]{mei18}
{Mei}, Z.~X., {Keppens}, R., {Roussev}, I.~I., \& {Lin}, J. 2018, \aap, 609, A2

\bibitem[{{Moore} {et~al.}(2001){Moore}, {Sterling}, {Hudson}, \&
  {Lemen}}]{2001ApJ...552..833M}
{Moore}, R.~L., {Sterling}, A.~C., {Hudson}, H.~S., \& {Lemen}, J.~R. 2001,
  \apj, 552, 833

\bibitem[{{Ning}(2016)}]{ning16}
{Ning}, Z. 2016, \apss, 361, 22

\bibitem[{Pesnell {et~al.}(2012)Pesnell, Thompson, \& Chamberlin}]{pesnel12}
Pesnell, W.~D., Thompson, B.~J., \& Chamberlin, P.~C. 2012, Solar Physics, 275,
  3

\bibitem[{Qin {et~al.}(2017)Qin, Na, Ju, \& Haimin}]{qin17}
Qin, L., Na, D., Ju, J., \& Haimin, W. 2017, The Astrophysical Journal, 841,
  112

\bibitem[{{Schou} {et~al.}(2012){Schou}, {Scherrer}, {Bush}, {Wachter},
  {Couvidat}, {Rabello-Soares}, {Bogart}, {Hoeksema}, {Liu}, {Duvall}, {Akin},
  {Allard}, {Miles}, {Rairden}, {Shine}, {Tarbell}, {Title}, {Wolfson},
  {Elmore}, {Norton}, \& {Tomczyk}}]{schou12}
{Schou}, J., {Scherrer}, P.~H., {Bush}, R.~I., {et~al.} 2012, \solphys, 275,
  229

\bibitem[{{Schrijver} {et~al.}(2008){Schrijver}, {Elmore}, {Kliem},
  {T{\"o}r{\"o}k}, \& {Title}}]{schr08}
{Schrijver}, C.~J., {Elmore}, C., {Kliem}, B., {T{\"o}r{\"o}k}, T., \& {Title},
  A.~M. 2008, \apj, 674, 586

\bibitem[{{Shen} \& {Liu}(2012)}]{2012ApJ...754....7S}
{Shen}, Y., \& {Liu}, Y. 2012, \apj, 754, 7

\bibitem[{{Shen} {et~al.}(2015){Shen}, {Liu}, {Liu}, {Chen}, {Su}, {Xu}, \&
  {Liu}}]{shen15}
{Shen}, Y., {Liu}, Y., {Liu}, Y.~D., {et~al.} 2015, \apjl, 814, L17

\bibitem[{{Shen} {et~al.}(2012{\natexlab{a}}){Shen}, {Liu}, \&
  {Su}}]{2012ApJ...750...12S}
{Shen}, Y., {Liu}, Y., \& {Su}, J. 2012{\natexlab{a}}, \apj, 750, 12

\bibitem[{{Shen} {et~al.}(2012{\natexlab{b}}){Shen}, {Liu}, {Su}, \&
  {Deng}}]{2012ApJ...745..164S}
{Shen}, Y., {Liu}, Y., {Su}, J., \& {Deng}, Y. 2012{\natexlab{b}}, \apj, 745,
  164

\bibitem[{{Shen} {et~al.}(2017{\natexlab{a}}){Shen}, {Liu}, {Tian}, \&
  {Qu}}]{2017ApJ...851..101S}
{Shen}, Y., {Liu}, Y., {Tian}, Z., \& {Qu}, Z. 2017{\natexlab{a}}, \apj, 851,
  101

\bibitem[{{Shen} {et~al.}(2017{\natexlab{b}}){Shen}, {Liu}, {Su}, {Qu}, \&
  {Tian}}]{2017ApJ...851...67S}
{Shen}, Y., {Liu}, Y.~D., {Su}, J., {Qu}, Z., \& {Tian}, Z. 2017{\natexlab{b}},
  \apj, 851, 67

\bibitem[{{Shen} {et~al.}(2011){Shen}, {Liu}, \& {Liu}}]{2011RAA....11..594S}
{Shen}, Y.-D., {Liu}, Y., \& {Liu}, R. 2011, Research in Astronomy and
  Astrophysics, 11, 594

\bibitem[{{Su} {et~al.}(2007){Su}, {Liu}, {Kurokawa}, {Mao}, {Yang}, {Zhang},
  \& {Wang}}]{su07}
{Su}, J., {Liu}, Y., {Kurokawa}, H., {et~al.} 2007, \solphys, 242, 53

\bibitem[{{Tian} {et~al.}(2017){Tian}, {Liu}, {Shen}, {Elmhamdi}, {Su}, {Liu},
  \& {Kordi}}]{tian17}
{Tian}, Z., {Liu}, Y., {Shen}, Y., {et~al.} 2017, \apj, 845, 94

\bibitem[{{T{\"o}r{\"o}k} {et~al.}(2004){T{\"o}r{\"o}k}, {Kliem}, \&
  {Titov}}]{toeroe04}
{T{\"o}r{\"o}k}, T., {Kliem}, B., \& {Titov}, V.~S. 2004, \aap, 413, L27

\bibitem[{{van Ballegooijen} \& {Martens}(1989)}]{van89}
{van Ballegooijen}, A.~A., \& {Martens}, P.~C.~H. 1989, \apj, 343, 971

\bibitem[{Wang \& Muglach(2007)}]{wang07}
Wang, Y.-M., \& Muglach, K. 2007, The Astrophysical Journal, 666, 1284

\bibitem[{Wang \& Muglach(2013)}]{wang13}
---. 2013, The Astrophysical Journal, 763

\bibitem[{{Xue} {et~al.}(2014){Xue}, {Yan}, {Qu}, \&
  {Zhao}}]{2014NewA...26...23X}
{Xue}, Z., {Yan}, X., {Qu}, Z., \& {Zhao}, L. 2014, \na, 26, 23

\bibitem[{{Yan} {et~al.}(2015){Yan}, {Xue}, {Pan}, {Wang}, {Xiang}, {Kong}, \&
  {Yang}}]{yan15}
{Yan}, X.~L., {Xue}, Z.~K., {Pan}, G.~M., {et~al.} 2015, \apjs, 219, 17

\bibitem[{{Yang} {et~al.}(2016{\natexlab{a}}){Yang}, {Jiang}, {Yang}, {Bi}, \&
  {Li}}]{2016ApJ...830...16Y}
{Yang}, B., {Jiang}, Y., {Yang}, J., {Bi}, Y., \& {Li}, H. 2016{\natexlab{a}},
  \apj, 830, 16

\bibitem[{{Yang} {et~al.}(2015){Yang}, {Jiang}, {Yang}, {Hong}, \&
  {Xu}}]{2015ApJ...803...86Y}
{Yang}, B., {Jiang}, Y., {Yang}, J., {Hong}, J., \& {Xu}, Z. 2015, \apj, 803,
  86

\bibitem[{{Yang} {et~al.}(2016{\natexlab{b}}){Yang}, {Jiang}, {Yang}, {Yu}, \&
  {Xu}}]{2016ApJ...816...41Y}
{Yang}, B., {Jiang}, Y., {Yang}, J., {Yu}, S., \& {Xu}, Z. 2016{\natexlab{b}},
  \apj, 816, 41

\bibitem[{{Yang} {et~al.}(2012{\natexlab{a}}){Yang}, {Jiang}, {Yang}, {Hong},
  {Yang}, {Bi}, {Zheng}, \& {Li}}]{2012NewA...17..732Y}
{Yang}, J., {Jiang}, Y., {Yang}, B., {et~al.} 2012{\natexlab{a}}, \na, 17, 732

\bibitem[{{Yang} {et~al.}(2012{\natexlab{b}}){Yang}, {Jiang}, {Yang}, {Zheng},
  {Yang}, {Hong}, {Li}, \& {Bi}}]{2012SoPh..279..115Y}
---. 2012{\natexlab{b}}, \solphys, 279, 115

\bibitem[{{Yang} {et~al.}(2012{\natexlab{c}}){Yang}, {Jiang}, {Zheng}, {Bi},
  {Hong}, \& {Yang}}]{2012ApJ...745....9Y}
{Yang}, J., {Jiang}, Y., {Zheng}, R., {et~al.} 2012{\natexlab{c}}, \apj, 745, 9

\bibitem[{{Yang} {et~al.}(2011){Yang}, {Jiang}, {Zheng}, {Hong}, {Bi}, \&
  {Yang}}]{2011SoPh..270..551Y}
---. 2011, \solphys, 270, 551

\bibitem[{{Yang} {et~al.}(2017){Yang}, {Yan}, {Li}, {Xue}, \&
  {Xiang}}]{2017ApJ...838..131Y}
{Yang}, L., {Yan}, X., {Li}, T., {Xue}, Z., \& {Xiang}, Y. 2017, \apj, 838, 131

\bibitem[{{Yang} {et~al.}(2013){Yang}, {Zhang}, {Liu}, {Li}, \&
  {Shen}}]{2013ApJ...775...39Y}
{Yang}, L., {Zhang}, J., {Liu}, W., {Li}, T., \& {Shen}, Y. 2013, \apj, 775, 39

\bibitem[{{Yeates} {et~al.}(2008){Yeates}, {Mackay}, \& {van
  Ballegooijen}}]{2008SoPh..247..103Y}
{Yeates}, A.~R., {Mackay}, D.~H., \& {van Ballegooijen}, A.~A. 2008, \solphys,
  247, 103

\bibitem[{{Yokoyama} \& {Shibata}(1995)}]{1995Natur.375...42Y}
{Yokoyama}, T., \& {Shibata}, K. 1995, \nat, 375, 42

\bibitem[{{Zhang} {et~al.}(2014){Zhang}, {Liu}, {Wang}, {Shen}, {Liu}, {Liu},
  \& {Wang}}]{zhang14}
{Zhang}, Q., {Liu}, R., {Wang}, Y., {et~al.} 2014, \apj, 789, 133

\bibitem[{{Zhang} {et~al.}(2017){Zhang}, {Li}, \& {Ning}}]{2017ApJ...851...47Z}
{Zhang}, Q.~M., {Li}, D., \& {Ning}, Z.~J. 2017, \apj, 851, 47

\bibitem[{Zhu \& Alexander(2014)}]{zhu14}
Zhu, C., \& Alexander, D. 2014, Solar Physics, 289, 279

\end{thebibliography}
\end{document}